\documentclass[aps,pra,reprint,showpacs,superscriptaddress,floatfix]{revtex4-1}
\usepackage[pdftex]{graphicx}
\usepackage{dcolumn}
\usepackage{bm}
\usepackage[usenames, dvipsnames]{color}
\usepackage{comment}

\usepackage[T1]{fontenc}

\usepackage{enumerate}
\usepackage{colonequals}
\usepackage{amsfonts}
\usepackage{amssymb}

\usepackage{amsthm}
\usepackage{bbm}
\usepackage{multirow}
\usepackage[colorinlistoftodos]{todonotes}
\usepackage[scr=boondoxo, scrscaled=1.05]{mathalfa}
\usepackage{amsmath}

\allowdisplaybreaks

\DeclareFontFamily{T1}{pzc}{}
\DeclareFontShape{T1}{pzc}{m}{it}{<-> [1.18] pzcmi8t}{}
\DeclareMathAlphabet{\mathpzc}{T1}{pzc}{m}{it}

\usepackage{hyperref}
\hypersetup{
     colorlinks   = true,
     citecolor    = blue,
     linkcolor  =black,
     urlcolor   =black}

\usepackage{subfigure}

\usepackage{letltxmacro}
\makeatletter
\let\oldr@@t\r@@t
\def\r@@t#1#2{%
\setbox0=\hbox{$\oldr@@t#1{#2\,}$}\dimen0=\ht0
\advance\dimen0-0.2\ht0
\setbox2=\hbox{\vrule height\ht0 depth -\dimen0}%
{\box0\lower0.4pt\box2}}
\LetLtxMacro{\oldsqrt}{\sqrt}
\renewcommand*{\sqrt}[2][\ ]{\oldsqrt[#1]{#2}}
\makeatother

\usepackage{array}
\newcolumntype{L}[1]{>{\raggedright\let\newline\\\arraybackslash\hspace{0pt}}m{#1}}
\newcolumntype{C}[1]{>{\centering\let\newline\\\arraybackslash\hspace{0pt}}m{#1}}
\newcolumntype{R}[1]{>{\raggedleft\let\newline\\\arraybackslash\hspace{0pt}}m{#1}}
\usepackage{tabularx}

\def\KeyWord#1{$\backslash$\IfColor{$\!\!$\textRed{#1}\textBlack}{#1}$\!\!$}

\newcommand{\be}{\begin{equation} }
\newcommand{\ee}{\end{equation} }
\newcommand{\ba}{\begin{eqnarray} }
\newcommand{\ea}{\end{eqnarray} }
\newcommand{\bit}{\begin{itemize}}
\newcommand{\eit}{\end{itemize}}
\newcommand{\ben}{\begin{enumerate}}
\newcommand{\een}{\end{enumerate}}

\newcommand{\x}{\sigma^{x}}
\newcommand{\y}{\sigma^{y}}
\newcommand{\z}{\sigma^{z}}

\newcommand{\m}{\sigma^{-}}

\renewcommand{\d}{\mathrm{d}}
\newcommand{\e}{\mathrm{e}}

\def\bra#1{\langle#1|}
\def\ket#1{|#1\rangle}

\def\cexp#1{\langle#1\rangle}

\def\e{\mathrm{e}}

\def\up{\uparrow}
\def\dn{\downarrow}

\def\n{{\vec{n}}}
\def\m{{\vec{m}}}

\begin{document}
\title{Mean field theory of failed thermalizing avalanches}

\author{P. J. D. Crowley}
\email{philip.jd.crowley@gmail.com}
\affiliation{Department of Physics, Massachusetts Institute of Technology, Cambridge, Massachusetts 02139, USA}

\author{A. Chandran}
\affiliation{Department of Physics, Boston University, Boston, Massachusetts 02215, USA}

\date{\today}

\begin{abstract}
We show that localization in quasiperiodically modulated, two-dimensional systems is stable to the presence of a finite density of ergodic grains. This contrasts with the case of randomly modulated systems, where such grains seed thermalizing avalanches. These results are obtained within a quantitatively accurate, self-consistent \emph{entanglement mean field theory} which analytically describes two level systems connected to a central ergodic grain. The theory predicts the distribution of entanglement entropies of each two level system across eigenstates, and the late time values of dynamical observables. In addition to recovering the known phenomenology of avalanches, the theory reproduces exact diagonalization data, and predicts the spatial profile of the thermalized region when the avalanche fails.
\end{abstract}

\maketitle

In the presence of sufficiently strong disorder, an interacting many-body quantum system may become \emph{many-body localized} (MBL)~\cite{fleishman1980interactions,gornyi2005interacting,basko2006metal,basko2006problem,oganesyan2007localization,pal2010many,luitz2015many,ponte2015many,imbrie2016many,imbrie2016diagonalization,abanin2017recent,alet2018many,abanin2019colloquium,nandkishore2015many,long2021many}. Local subsystems of an MBL system do not thermalize, and instead retain memory of their initial conditions indefinitely. MBL systems thus provide remarkable counterexamples to the ergodic hypothesis, and are outside the scope of quantum statistical mechanics~\cite{nandkishore2015many,altman2018many}. Instead, the phenomenology of these systems is dictated by an extensive set of emergent and exponentially localized conserved operators, known as \emph{local integrals of motion} or \emph{l-bits}~\cite{huse2014phenomenology,serbyn2013local,ros2015integrals,chandran2015constructing,pekker2017fixed,imbrie2017local}.

Even when typical regions appear strongly localized, thermalizing avalanches seeded by rare ergodic inclusions may destabilize MBL~\cite{de2017stability}. This instability places restrictions on the existence of MBL. Consider an \emph{ergodic grain}---a microscopically small spatial region which locally thermalizes. The grain can serve as a bath, thermalizing nearby l-bits. These thermalized l-bits are absorbed into the bath, forming a \emph{thermal bubble}. The density of states of the thermal bubble is enhanced, increasing exponentially in the number $n$ of l-bits absorbed. However, the coupling strength of an l-bit to the bubble decays exponentially in its spatial separation $r$ from the grain. In dimension $D$ the separation of the $n$th most strongly coupled l-bit scales as $r \sim n^{1/D}$. Which of these two effects dominates depends on $D$.

In $D=1$ the enhancement to the bubble and the smallness of the couplings are both exponential in $n$, leading to a competition~\cite{de2017stability,de2017many,luitz2017small,thiery2018many,goremykina2019analytically,dumitrescu2019kosterlitz,crowley2020avalanche,sels2021markovian,morningstar2021avalanches}. If the exponential decay constant of the couplings is slower than a critical rate, the enhancement prevails, and an \emph{avalanche} occurs in which all l-bits are absorbed into the thermal bubble. In contrast, if the decay of the couplings is sufficiently fast, MBL is stable as the avalanche halts after absorbing finitely many l-bits. If, in the thermodynamic limit, the number of thermalized l-bits is large, but sub-extensive, the grain may be regarded as having induced a \emph{failed avalanche}.

In $D > 1$ the enhancement of the bubble always prevails asymptotically~\cite{de2017stability,gopalakrishnan2019instability,potirniche2019exploration}. Whether avalanches occur instead depends only on whether the microscopic environment of the grain allows the avalanche to reach this asymptotic regime. However, such effects of the microscopic environment may be overcome by larger initial grains. For thermodynamic systems with uncorrelated random disorder, arbitrarily large ergodic grains occur in the system. Thus, grains large enough to seed avalanches necessarily exist with a finite density, destabilizing the putative MBL phase. In contrast, it is believed that if the localizing spatial potential is highly correlated (e.g. with quasiperiodic modulation), there may be no ergodic grains of sufficient size to start an avalanche. That is, the correlated potential can cause \emph{all} putative avalanches to fail at a finite size, allowing for stable MBL in $D>1$.

However, a quantitative theory of how avalanches fail is lacking. For example, it is not known how strongly coupled to the bubble an l-bit must be to thermalize; how to treat groups of l-bits with comparable couplings to the bubble; or how to account for l-bits that are only partially thermalized. Moreover, the presence of a collar of partially thermalized l-bits blurs the bubble's boundaries, making it unclear how to quantify its size~\cite{crowley2020avalanche}. 

\begin{figure}[t!]
    \centering
    \includegraphics[width=\linewidth]{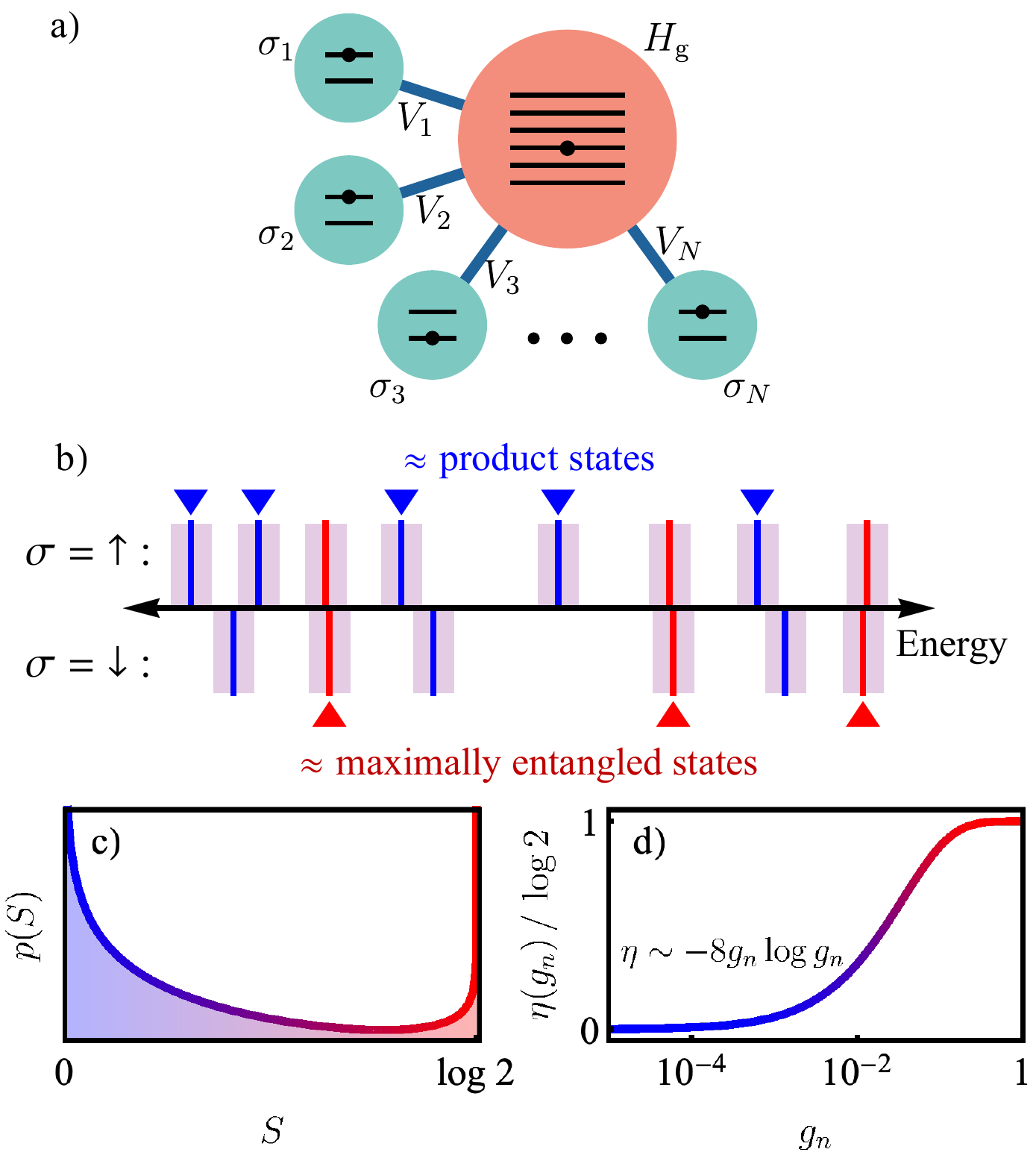}
    \caption{
    \emph{Non ergodicity in central grain models:} a) The central grain model~\eqref{eq:H_central_spin} consists of two level systems coupled to a grain. The coupling operators $V_n$ may vary independently. b) For $V_n = 0$ the spectrum can be divide into two sectors: $\sigma_n = \up/\dn$. When $V_n \neq 0$ is in the intermediate regime, most eigenstates remain close to product states of the l-bit and grain (blue). However, a minority of the $V_n = 0$ eigenstates are close to a state in the opposite sector, and consequently form resonances when $V_n$ is introduced (red). The relevant scale of closeness is set by the matrix elements (purple collars). c) In the intermediate regime the distribution of eigenstate entanglement entropies of $\sigma_n$ over eigenstates, $p_{S_{na}}(S)$, is bi-modal: approximate product states contribute the mode at $S = 0$, and resonances contribute the mode at $S = \log 2$. d) The formation of resonances with $\sigma_n$ enhances the entropy of the grain by $\eta(g_n)$, where $g_n$ is the reduced coupling~\eqref{eq:g1}. 
    }
    \label{Fig:1}
\end{figure}

In this manuscript, we develop a quantitative theory of failed avalanches in a toy model. Specifically, we develop an \emph{entanglement mean field theory} of l-bits coupled to a central thermalizing grain (Fig~\ref{Fig:1}a). This theory captures the enhancement of the bubble due to partially thermalized l-bits. Quantitatively, it predicts the distribution of l-bit entanglement entropies across eigenstates, and late time values of l-bit observables in dynamical experiments.
We apply this theory to study avalanches in $D=1$ and $D=2$. In $D=1$, the theory exhibits quantitative agreement with exact diagonalization. In $D=2$, for a single finite grain and sufficiently strong quasiperiodic modulation, avalanches always fail, and the number of l-bits in the thermal bubble is bounded. Using this bound we establish that a finite density of regularly spaced grains does not induce an avalanche. In contrast, for arbitrarily strong random modulation, we find that a single grain has a non-zero probability of inducing an avalanche.

\paragraph*{\textbf{Central grain model}:} We consider l-bits, here spins-$1/2$, which are coupled to a central few level system, or grain (Fig.~\ref{Fig:1}a)
\begin{equation}
    H = H_\mathrm{g} + \sum_{n=1}^N h_n \z_n + \sum_{n=1}^N V_n
    \label{eq:H_central_spin}
\end{equation}
where $H_\mathrm{g}$ is the grain Hamiltonian, $h_n$ sets the splitting of the $n$th l-bit, and the coupling operator $V_n$ both flips the $n$th spin, and acts non-trivially on the grain $[V_n,H_\mathrm{g}]\neq 0$, e.g.
\begin{equation}
    V_n = J_n (\x_\mathrm{g} \x_n + \y_\mathrm{g} \y_n).
    \label{eq:coupling}
\end{equation}
Here $\sigma_\mathrm{g}^\alpha$ acts on the grain, and the $J_n$ are coupling constants. In toy models of the localized phase, the $J_n$ are typically exponentially decaying in $n$. We assume the grain has a density of states at maximum entropy $\rho_\mathrm{g}$ and is ergodic. In technical terms: physical operators on the grain satisfy the off-diagonal eigenstate thermalization hypothesis (ETH) in the eigenbasis of $H_\mathrm{g}$, see Refs.~\cite{jensen1985statistical,deutsch1991quantum,srednicki1994chaos,rigol2008thermalization,kim2014testing,d2016quantum,brenes2020low,beugeling2015off}. For simplicity we restrict to the case where the splittings $h_n$ are below the bandwidth of the grain $h_n \ll |H_\mathrm{g}|$, but above its energy level spacing $h_n \rho_\mathrm{g} \gg 1$. Central grain models have been previously studied, both in their own right~\cite{ponte2017thermal,hetterich2018detection,ashida2019quantum}, and as toy models for ergodic inclusions in the MBL phase~\cite{de2017stability,luitz2017small,potirniche2019exploration,crowley2020avalanche}.

\paragraph*{\textbf{Single spin case}:} We begin with the simplest case, that of $N=1$, where we recap the relevant parts of Ref.~\cite{crowley2022partial} in which this problem was analysed in detail. The strength of the spin-grain coupling is characterised by a single dimensionless quantity: the reduced coupling $g_1$. The reduced coupling is defined as the mean off-diagonal matrix element of the coupling operator $V_1$ measured in units of the level spacing
\begin{equation}
    g_1 : = \rho \, [|V_{1,ab}|], \qquad V_{1,ab} = \bra{E_a} V_1 \ket{E_b}
    \label{eq:g1}
\end{equation}
where $[\cdot ]$ denotes the mean over the indices $a \neq b$, $\rho$ and $\ket{E_a}$ are respectively the density of states at maximum entropy and eigenbasis of $H$, both evaluated for $V_1 = 0$. The reduced coupling is given in terms of macroscopic quantities by
\begin{equation}
    g_1 = g_{1,0} : = \sqrt{2 \rho_\mathrm{g} v_1(h_1)/\pi }
    \label{eq:gn_spec_fun}
\end{equation}
where $v_1(\omega)$ is the infinite temperature spectral function of $V_1$ when evolved under $\left. H\right|_{V_1 = 0}$~\footnote{We note the numerical constants in~\eqref{eq:gn_spec_fun} are accurate for real $H_\mathrm{g}$, and are altered for complex or quaternionic $H_\mathrm{g}$, see Ref.~\cite{crowley2022partial}.}. 

The reduced coupling determines the sensitivity of the eigenstates to switching on $V_1$. Specifically, let $g_{1a}$ denote the $L_2$ norm of the first order term in perturbation theory when the eigenstate $\ket{E_a}$ is expanded about $V_1 = 0$
\begin{equation}
    g_{1a} = \sqrt{\sum_{b \neq a}  \left| \frac{V_{1,ab}}{E_a - E_b} \right|^2}.
    \label{eq:g1a}
\end{equation}
The norm $g_{1a}$ follows a distribution $p_{g_{1a}}(g_{1a})$ which may be exactly calculated. Typical values drawn from this distribution are on the scale of the reduced coupling $[g_{1a}]_\mathrm{typ.} = c g_1$ (for an $O(1)$ numerical constant $c$), however, the heavy power law tail
\begin{equation}
    p_{g_{1a}} \sim 2 g_1/g_{1a}^2
    \label{eq:pg1a}
\end{equation}
implies the frequent occurrence of much larger values. This tail is a generic and robust feature which is due to \emph{resonances}: pairs of states which are accidentally close in energy, and so strongly hybridize upon even very weak perturbations (Fig.~\ref{Fig:1}b).

The reduced coupling dictates three distinct regimes of eigenstate structure, which manifest in corresponding regimes of late time dynamical behaviour. We discuss these regimes in turn:

\paragraph*{Strong coupling $(g_1 \gtrsim 1)$:} In this regime, typical eigenstates are non-perturbatively corrected by the coupling. This results in strong mixing between eigenstates, and the eigenstates of the combined system of spin and grain satisfy ETH. Specifically: the entanglement entropy of the spin $S_{1a}$ in the state $\ket{E_a}$, is close to maximal value of $\log 2$ for all states close to maximum entropy, and the spin has no infinite time memory of its initial condition, as characterised by the infinite temperature time-averaged spin-spin correlator $\overline{C_1^{zz}} = \overline{\cexp{\z_1(t)\z_1(0)}}$:
\begin{subequations}
\begin{equation}
    [S_{1a}]  = \log 2, \quad 
    \overline{C_1^{zz}} = 0,
    \quad \text{(strong coupling)}.
\end{equation}
where here $[\cdot]$ denotes the average over states at maximum entropy.
For brevity, we here neglect corrections in $g_1^{-1}$, and throughout we neglect finite size corrections which are small in $1/d_\mathrm{g}$ where $d_\mathrm{g} := \dim(H_\mathrm{g})$ is the grain dimension.

\paragraph*{Weak coupling $(g_1 \ll 1/d_\mathrm{g})$:} For weak coupling, in a typical realization $\max_a g_{1a} \ll 1$, so that all the eigenstates, including those in the tail of $p_{g_{1a}}$ experience only perturbative corrections from the zero coupling ($V_1 = 0$) limit of $[S_a]=0$ and $\overline{C_1^{zz}} = 1$.

\paragraph*{Intermediate coupling $(1 \gg g_1 \gtrsim 1/d_\mathrm{g})$:} In the intervening regime typical eigenstates are only perturbatively corrected (as for weak coupling), whereas an $O(g_1)$ fraction of states are rare resonances, which have $g_{1a}>1$. The resonant eigenstates are non-perturbatively corrected, and consequently attain large spin entanglement entropies $S \approx \log 2$. In the intermediate regime, the distribution of eigenstate entanglement entropies is bi-modal (Fig.~\ref{Fig:1}c), with the resonances forming the dominant contribution to the mean
\begin{equation}
    [S_{1a}]  = 2 \pi g_1 \log 2, \quad 
    \overline{C_1^{zz}} = 1 - k g_1,
    \quad \text{(int. coupling)},
    \label{eq:S_Czz}
\end{equation}
\end{subequations}
where we have neglected subleading $O(g_1^2)$ corrections, and the constant $k$ may be calculated. We emphasise: though resonances are identified using $g_{1a}$, a quantity that is perturbative in nature, the subsequent treatment is non-perturbative. Consequently,~\eqref{eq:S_Czz} remains accurate throughout the intermediate regime, up to the crossover to strong coupling $g_1 \gtrsim 1$. See App.~\ref{app:inf_time} for precise forms of $[S_{1a}]$, $\overline{C_1^{zz}}$ and the distribution of entanglement entropies over eigenstates $p_{S_{1a}}(S_{1a})$ accurate for all $g_1$, and Ref.~\cite{crowley2022partial} for an explanation of these results.

\paragraph*{\textbf{Two spin case}:} We now consider the $N=2$ case. Suppose first, that the second spin is in the weak coupling regime irrespective of the value of $g_1$. The coupling strength of the second spin is characterised by a reduced coupling $g_2$, defined analogously to $g_1$~\eqref{eq:g1}. However, as the second spin sees an effective thermal bath comprising the first spin and central grain, $g_2$ must be defined using the eigenbasis and density of states calculated for the combined system of the first spin and central grain. When $g_1$ is in the weak coupling regime, we thus obtain
\begin{subequations}
\begin{equation}
    g_2 = g_{2,0} := \sqrt{2 \rho_\mathrm{g} v_2(h_2)/\pi }, \qquad (g_1 \text{ weak}).
\end{equation}
in direct correspondence with~\eqref{eq:gn_spec_fun}. In contrast, if $g_1$ is strongly coupled, $g_2$ is enhanced~\cite{de2017stability,crowley2022partial}
\begin{equation}
    g_2 = \sqrt{2} g_{2,0}, \qquad \quad (g_1 \text{ strong}).
\end{equation}
In the intermediate regime, the reduced coupling $g_2$ is intermediately enhanced. This intermediate enhancement to $[|V_{2,ab}|]$ may be calculated by accounting for resonances (see App.~\ref{app:eta})
\begin{equation}
    g_2 = g_{2,0} \e^{ \eta(g_1) / 2}, \qquad 0 \leq \eta(g_1) \leq \log 2
    \label{eq:g2_int}
\end{equation}
\end{subequations}
Here $\eta(g_1)$ is the entropic enhancement to the central grain due to the resonances formed upon coupling to the first spin. This function smoothly interpolates between the small and large limits of $\eta(g) \sim -8 g \log g $ and $\eta(g) = \log 2 + O(g^{-4})$. An analytic form for $\eta(g)$ is calculated and verified in Ref.~\cite{crowley2022partial}, quoted in App.~\ref{app:eta}, and plotted in Fig.~\ref{Fig:1}d.

We briefly comment on how~\eqref{eq:g2_int} should be quantitatively understood. As before, we may characterise the effect of coupling to the second spin by calculating $p_{g_{2a}}$, the distribution of $g_{2a}$. Here $g_{2a}$ is the $L_2$ norm of the first order correction in perturbation theory to the eigenstate $\ket{E_a}$ upon introducing the coupling $V_2$, \emph{but with $V_1$ finite}. For the $g_1$ intermediate regime, $p_{g_{2a}}$ will differ from $p_{g_{1a}}$ in details, however it has an identical power-law tail of non-perturbatively corrected states, i.e. resonances, $p_{g_{2a}} \sim 2 g_2/g_{2a}^2$, with $g_2$ given by~\eqref{eq:g2_int} (see App.~\ref{app:eta}). As before, this tail of resonances dictates eigenstate entanglement entropy, and long time memory, via~\eqref{eq:S_Czz} (with the index changed as appropriate).

When the first \emph{and} second spins are in the intermediate regime, the second spin sees an effective bath which is enhanced as compared to the bare grain due to hybridization with the first spin, and \emph{vice versa}. As a result the entropic enhancement must be solved self-consistently
\begin{equation}
    g_1 = g_{1,0} \e^{ \eta(g_2) / 2}, \quad g_2 = g_{2,0} \e^{ \eta(g_1) / 2}
\end{equation}
This results in a pair of solutions whose reduced couplings are enhanced over the bare seed properties $g_1 \geq g_{1,0}$, $g_2 \geq g_{2,0}$. We note that the second spin sees an effective bath which is enhanced to a greater degree than might naively be expected by hybridization between the first spin and grain alone $g_2 \geq g_{2,0} \e^{ \eta(g_{1,0}) / 2} $. Physically, this additional enhancement originates with the formation of resonances involving both spins in addition to the resonances involving the grain and one or other of the spins.

\paragraph*{\textbf{Generic ($N$-spin) case}:} We now introduce a third spin which is weakly coupled, and thus does not enhance the effective bath. Its coupling to the grain is characterised by the reduced coupling $g_3$. Remarkably, the entropic enhancement to the grain due to hybridizing with the first and second spin takes a simple additive form (see App.~\ref{app:eta})
\begin{equation}
    g_3 = g_{3,0} \e^{ (\eta(g_1) + \eta(g_2))/2}.
\end{equation}
This may be further generalized to a self-consistency equation for the many-spin case
\begin{equation}
    g_n = g_{n,0} \exp \bigg( \tfrac12 \sum_{m \neq n} \eta(g_m) \bigg).
    \label{eq:MFeqs_N}
\end{equation}
Several comments are in order. Firstly we emphasise that the self-consistency equations constitute a mean-field-like approximation: specifically we characterise the distribution $p_{g_n}$ of the $L_2$ norms $g_{na}$ between each spin and the enhanced grain by a single value, $g_n$, which characterises the heavy tail $p_{g_n} \sim 2 g_n / g_{na}^2$. Secondly, we note that we do not assume that the effective bath (comprising the central grain and spins with which it is entangled) satisfies ETH. On the contrary, whenever spins are in the intermediate coupling regime, the effective bath comprising the central grain and spins exhibits marked deviation from ETH. This deviation from ETH may be regarded as an accurate accounting of the ``back action'' of the intermediately coupled spins onto the bubble. This is shown for the case $N=1$ in Ref.~\cite{crowley2022partial}, where the distribution of off diagonal elements of a local operator on the grain is shown to be highly non-Gaussian. The same result may be obtained for $N>1$ by direct generalization. Thus, our technique goes beyond approximations standard in the literature~\cite{de2017stability,crowley2020avalanche}, and quantifies the formation of entanglement in non-ETH central grain systems which are too large to be studied directly using exact diagonalization.

\begin{figure}[t!]
    \centering
    \includegraphics[width=\linewidth]{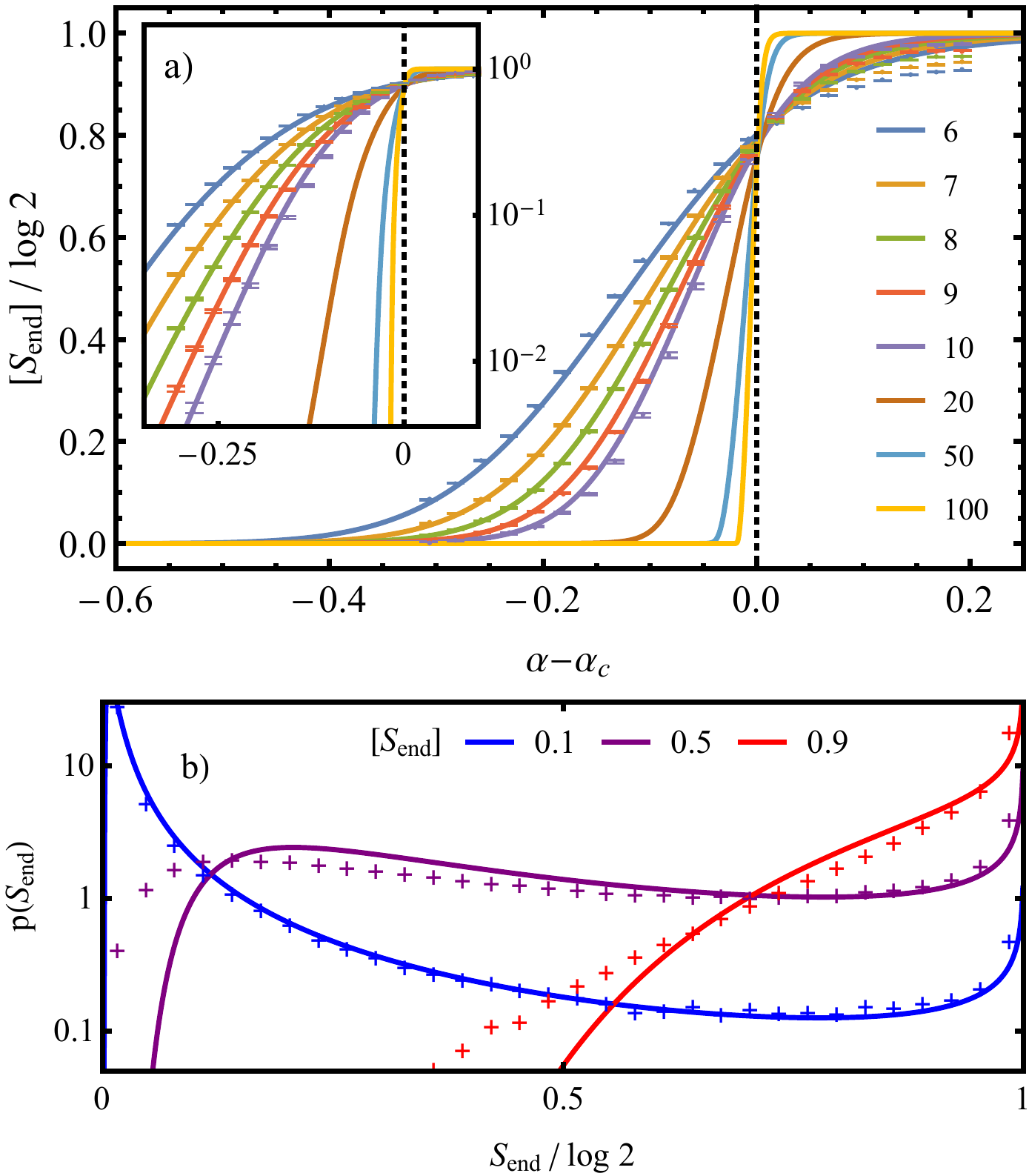}
    \caption{
    \emph{Comparison between the mean field equations~\eqref{eq:MFeqs_N} and exact diagonalization (ED) in the DRH model:} a) the state and sample averaged eigenstate entanglement entropy of the end l-bit $[S_\mathrm{end}]$ calculated via exact diagonalization for the DRH model is plotted (points with error bars), number of l-bits $N$ given in legend. The solid lines show the analytic mean field calculation. Inset: same data shown on a log scale. b): the sample averaged distribution over eigenstates $p(S_\mathrm{end})$ is shown for ED (points) and mean field calculation (solid) for different values of $\alpha$ (values of $[S_\mathrm{end}]$ given in legend to 2.d.p, and corresponding respectively to $\alpha = 0.45,0.55,0.65$). Other parameters in text.
    }
    \label{Fig:DRH1}
\end{figure}

\begin{figure}[t!]
    \centering
    \includegraphics[width=\linewidth]{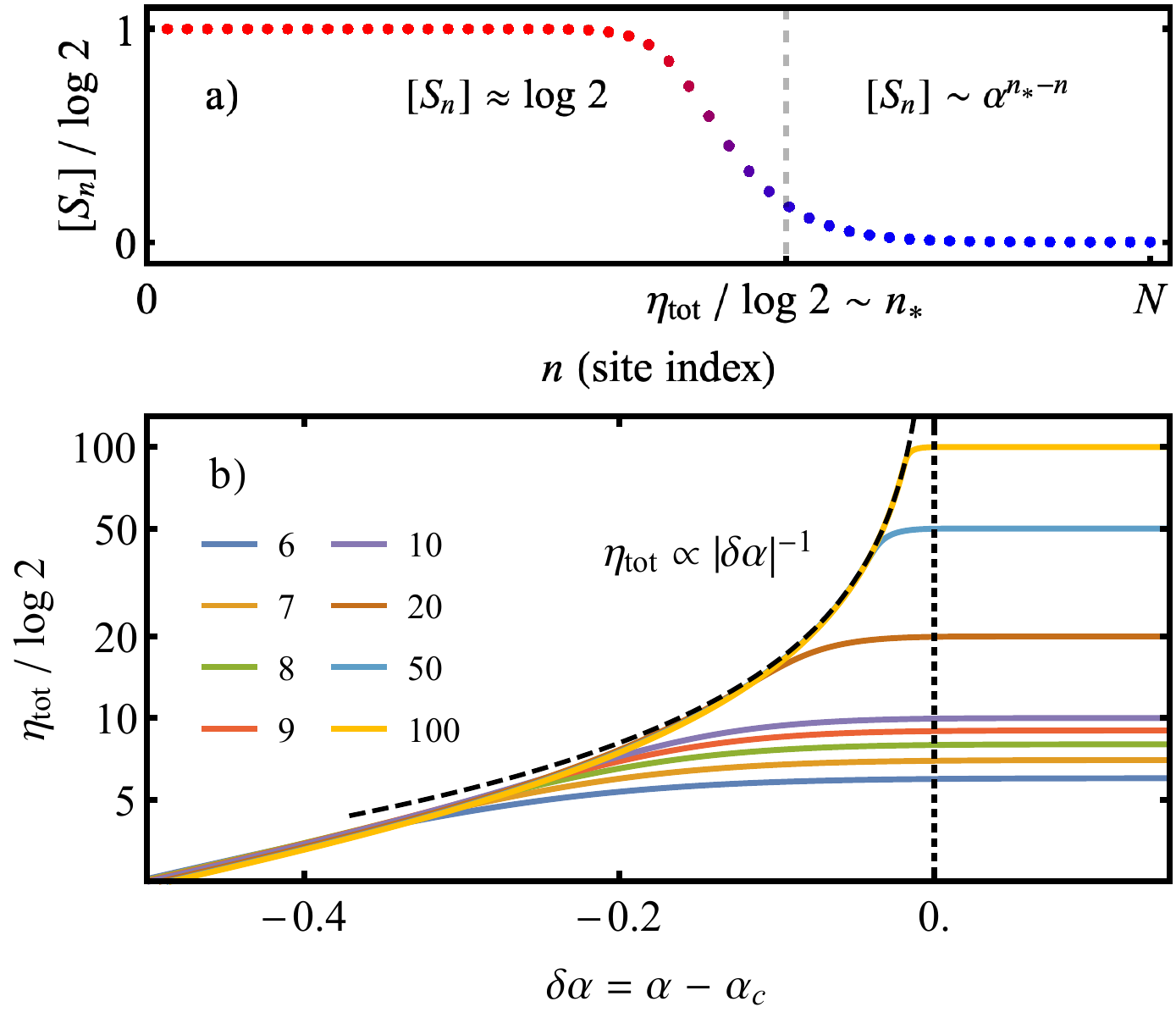}
    \caption{
    \emph{Failed avalanches in the DRH model:} a) The mean entanglement entropy $[S_n]$ vs site index $n$, for small, negative $\delta\alpha$ calculated using the mean field equations. The putative avalanche absorbs many ($n_\star = |\delta \alpha|^{-1}$) l-bits before failing. Parameters: $\delta \alpha = -0.07$, $N=50$, $g_{1,0} = 1/2$. b) The total entropic enhancement of the bubble $\eta_\mathrm{tot} \sim n_\star \log 2$ is shown for different system sizes $N$ (legend inset). For localized systems $\eta_\mathrm{tot} = O(N^0)$, for thermal systems $\eta_\mathrm{tot} = O(N^1)$. For sufficiently small $|\delta\alpha|$, and sufficiently large $N$, the critical scaling $\eta_\mathrm{tot} \propto |\delta \alpha|^{-1}$ (black dashed) emerges, corresponding to very large failed avalanches.
    }
    \label{Fig:DRH2}
\end{figure}

\paragraph*{\textbf{Avalanches in 1D}:}

The De Roeck and Huveneers (DRH) model~\cite{de2017stability,luitz2017small} is a minimal model of the avalanche instability of the MBL phase in $D=1$. Specifically, it corresponds to the model~\eqref{eq:H_central_spin} with exponentially decaying couplings between the grain and l-bits. We use the couplings~\eqref{eq:coupling} with 
\begin{equation}
    J_n = J_1 \alpha^{n-1}
    \label{eq:DRH_couplings}
\end{equation}
corresponding to a grain coupled to one end of a localized chain~\cite{luitz2017small}. The DRH model avalanches when the couplings decay slower than the critical value $\alpha > \alpha_\mathrm{c} = 1/\sqrt{2}$ and $J_1$ not pathologically large or small. For $\alpha < \alpha_\mathrm{c}$, the avalanche fails, and the thermal bubble absorbs a finite number $n_\star$ of l-bits. Close to the critical point this number diverges as $n_\star \sim | \alpha - \alpha_\mathrm{c}|^{-\nu}$ with $\nu = 1$~\cite{de2017stability,luitz2017small,crowley2020avalanche,vsuntajs2022ergodicity}.

We compare the predictions of mean-field theory~\eqref{eq:MFeqs_N} with exact diagonalization (ED) data from the DRH model. We show the theory provides a quantitatively accurate description of failed avalanches, including their spatial extent, and both the infinite time memory and eigenstate entanglement entropies of the putative l-bits.

The ED results are obtained for the DRH model, with the $h_n$ drawn uniformly from the interval $h_n \in h + [ -\delta h, \delta h ]$ and $H_\mathrm{g}$ a GOE matrix. We use parameters $h=1$, $\delta h = 0.1$, $J_1 = 0.41$, and $d_\mathrm{g} = \mathrm{dim}(H_\mathrm{g}) = 8$. The bandwidth of $H_\mathrm{g}$ is determined by the RMS eigenvalue $([\mathrm{tr}(H_\mathrm{g}^2)]/d_\mathrm{g})^{1/2} = 1.5$. The mean field equations are solved by iteration
\begin{equation}
    g_n = \lim_{k \to \infty } g_{n,k} ,\quad g_{n,k+1} = g_{n,0} \exp \bigg( \tfrac12 \sum_{m \neq n} \eta(g_{m,k}) \bigg)
\end{equation}
for parameters $g_{n,0} = g_{1,0} \alpha^{n-1}$. The mean and distribution of entanglement entropies are then extracted from the $g_n$ using the forms in App.~\ref{app:inf_time}, from Ref.~\cite{crowley2022partial}. As we are interested in verifying the mean field equations, and not the accuracy of our calculation of $g_{1,0}$, we fit $g_{1,0}$ so that the mean field and ED results agree at $\alpha = \alpha_\mathrm{c}$. 

The mean field equations quantitatively reproduce the ED results on the localized side. In Fig.~\ref{Fig:DRH1}a we plot the eigenstate and sample averaged entanglement entropy of the end (i.e. $n = N$) l-bit $[S_\mathrm{end}]$ as a function of the tuning parameter $\alpha$ extracted both from ED (points) and the mean field equations (solid curves). The mean field and ED data both display the same key features: for $\alpha > \alpha_\mathrm{c}$ the system avalanches thermalizing all $N$ l-bits, yielding $[S_\mathrm{end}] = \log 2$ up to finite size corrections. On the localized side, the avalanche fails before reaching the final l-bit yielding $[S_\mathrm{end}] = 0$, again up to finite size corrections. The crossover between these two limits extends over a range of $\alpha$ values of width $O(N^{-1})$, sharpening to a step at large $N$. 

For $\alpha < \alpha_\mathrm{c}$, and for all system sizes $N$, the mean field theory shows good quantitative agreement with ED. This may be contrasted with the noticeable discrepancy found on the thermal side for small $N$. Discrepancy on the thermal side may be expected, as the two-level resonance picture underlying the mean field equations becomes inaccurate in this regime.

In the lower panel, Fig~\ref{Fig:DRH1}b, we compare distributions of entanglement entropies over eigenstates: the ED data (points) is calculated for parameters $N = 10$, $\alpha = 0.45,0.55,0.65$ where $[S_\mathrm{end}] = 0.10, 0.50, 0.90$ (2 d.p.), and may be compared with the mean field theory (solid lines) corresponding to the same values of $[S_\mathrm{end}]$ (these correspond to slightly different values of $\alpha$ due to small discrepancies between theory and numerics in Fig~\ref{Fig:DRH1}a). We again find the mean field equations show excellent agreement on the localized side (i.e. for $[S_\mathrm{end}] = 0.1$) with visible discrepancies when $[S_\mathrm{end}]$ is larger.

By accurately accounting for the partial entropic enhancement to the central grain from l-bits in the intermediate coupling regime, the mean field equations describe the physics of failed avalanches. An example is shown in Fig.~\ref{Fig:DRH2}a where the mean field equations are solved for $N=50$ and $\delta\alpha := \alpha - \alpha_\mathrm{c} = -0.07$. The avalanche thermalizes the first $n_\star \approx 30$ l-bits, before failing. This leaves the subsequent $N - n_\star \approx 20$ l-bits with entanglement entropies which are exponentially decaying in $n$. The avalanche proceeds to this extent despite the modest scale of the initial reduced coupling $g_{1,0} = 1/2$. In general, as the critical point $\alpha_\mathrm{c}$ is approached, the avalanche fails after thermalizing a number $n_\star = O( |\delta \alpha|^{-1})$ l-bits. This behaviour can be seen by analysing the total entropic enhancement of the effective thermal bubble
\begin{equation}
    \eta_\mathrm{tot} = \sum_{n} \eta(g_n).
    \label{eq:eta_tot}
\end{equation}
Unlike $[S_\mathrm{end}]$, $\eta_\mathrm{tot}$ provides information on the spatial extent of a failed avalanche: on the localized side $\eta_\mathrm{tot}$ grows proportional to the number of thermalized l-bits $\eta_\mathrm{tot} \sim n_\star \log 2 = O( |\delta \alpha|^{-1})$, whereas for avalanched systems $\eta_\mathrm{tot} = N \log 2$. This $\nu = 1$ scaling is visible in Fig~\ref{Fig:DRH2}b where the mean field values of $\eta_\mathrm{tot}$ are plotted for different system sizes $N$.

\paragraph*{\textbf{Avalanches in 2D:}} 
We apply the mean field equations to understand failed avalanches in 2D systems, revealing the marked stability of quasiperiodically modulated systems in higher dimensions to avalanches.

MBL due to uncorrelated random disorder is not stable in dimensions $D>1$ as thermal grains sufficiently large to cause avalanches always occur~\cite{de2017stability,potirniche2019exploration}. Instead, MBL is stable only if the localizing potential is sufficiently correlated that all putative avalanches deterministically fail at small sizes, before the feedback argument would allow them to self sustain. It is believed that this may occur in systems with quasiperiodic (QP) modulation.

As a model for avalanches in higher dimensions we consider a grain coupled to a system of free fermions on a 2D square lattice
\begin{equation}
    \begin{aligned}
    H & = H_\mathrm{g} + H_\mathrm{2D} + c_\mathrm{g}^\dagger c_{\vec{0}} + c_{\vec{0}}^\dagger c_\mathrm{g}
    \\
    H_\mathrm{2D} & = \sum_{\mathrm{NN}} c_\n^\dagger c_\m + \sum_\n V(\theta_\n) c_\n^\dagger c_\n
    \end{aligned}
    \label{eq:2D_cgm}
\end{equation}
where the hopping is nearest neighbour only, and $c_\mathrm{g}^\dagger$ acts on the grain, which is coupled only to the $\vec{n} = \vec{0}$ lattice site. We compare two cases in which the potential is obtained by sampling the periodic function $V(\theta) = V(\theta + 2 \pi)$ either (i) quasi-periodically, in which case $\theta_\n = \vec{q} \cdot \n + \theta_0$ with $\vec{q} = (q_1,q_2) = \pi ( 1 + \sqrt{5}, 1 + \sqrt{3})$ or (ii) randomly, in which case the $\theta_\n$ are drawn independently and uniformly from the circle $\theta_\n \in [0,2 \pi]$. For the periodic function $V(\theta)$ we use an asymmetric triangular wave of amplitude $W$ obtained by linearly interpolating between the points $V(0) = W$, $V(2 \pi/q_1) = -W$, $V(2 \pi) = W$. This model has desirable simplicity: in the random case the on-site potentials are uncorrelated and follow a box distribution. In the QP case, the modulation ensemble has a single parameter, $\theta_0$, and, as $V(\theta)$ does not have an inversion centre, the resulting lattice does not have points of `almost inversion symmetry' as present in e.g. the Aubry-Andre model~\footnote{such almost inversion centres do not alter the universal components of the physics, but complicate analysis by causing certain regions to be markedly less localized (as measured by e.g. the orbital IPR in the lattice basis)}. 

The model~\eqref{eq:2D_cgm} may be brought to the central grain form~\eqref{eq:H_central_spin} by working in the diagonal basis $f_\n = \sum_\m \phi_{\n\m}c_\m$ of $H_\mathrm{2D}$
\begin{equation}
    H = H_\mathrm{g} + \sum_\n \epsilon_\n f_\n^\dagger f_\n + \sum_\n \phi_{\n\vec{0}} (c_\mathrm{g}^\dagger f_{\n} + f_{\n}^\dagger c_\mathrm{g} ).
    \label{eq:2D_cgm_diagform}
\end{equation}
The diagonal orbitals $f_\n$ are labelled with the index corresponding to the physical site $c_\n$ with greatest overlap. Specifically, we maximise the quantity $\prod_{\n}|\phi_{\n\n}|$ over permutations of the rows of $\phi_{\m\n}$~\footnote{such a labelling may be found by e.g. the Blossom algorithm}. We extract the diagonal orbitals numerically, for which it is necessary to use a finite lattice. We truncate to a finite lattice radius $R$ around the site $\n = \vec{0}$ (i.e. we keep only sites $\vec{n} =(n_1,n_2)$ satisfying $|n_1|+|n_2| \leq R$), yielding a number of l-bits $N = 2 R^2 + 2 R + 1$. We note that in this model the two level systems are fermionic orbitals, differing from~\eqref{eq:H_central_spin} where we considered spins, however this detail is unimportant and both the mean field equations, and avalanche phenomenology are unaltered. 

For sufficiently strong potential strength $W$, for both QP and random potentials, the avalanche may fail. To see why, consider an avalanche which has thermalized $(n-1)$-l-bits, we thus approximate $\eta(g_m) \approx 0 \,(\log 2)$ for $m \geq n$ ($m < n$) where we have ordered the l-bits by coupling strength. The coupling $J_n \approx J_0 \e^{-r_n/\zeta}$ to the next most strongly coupled l-bit is exponentially small in the distance $r_n \sim \sqrt{n}$, yielding a reduced coupling to the $n$th spin of
\begin{equation}
    \begin{aligned}
    g_{n} \propto J_n \exp\! \Big( \tfrac12 \!\sum_{m \neq n} \eta(g_m) \Big) \approx J_0  \e^{- \sqrt{ n} / \zeta+ (n-1) \log 2/2}.
    \end{aligned}
    \label{eq:gn_2d_aval}
\end{equation}
Asymptotically the quantity~\eqref{eq:gn_2d_aval} is increasing in $n$, implying that $g_{n} \gtrsim 1$ for all $n$ sufficiently large, leading to a self-sustaining avalanche. However, at smaller $n$, the avalanche must pass through a bottleneck corresponding to the minimum of~\eqref{eq:gn_2d_aval} over $n$. If, at this minimum, the reduced coupling to the next l-bit is not strong $\min_n g_n \ll 1$ then the naive avalanche argument indicates the avalanche ceases at (or before) reaching this size. This is the basic picture of avalanche failure. The mean field analysis here refines this argument in two ways. First, it quantitatively describes the avalanche. Second, it includes the previously missing physics, that sufficiently many l-bits with comparable weak coupling $g_n$ may provide the entropic enhancement necessary for the avalanche to continue. 

\begin{figure}[t!]
    \centering
    \includegraphics[width=\linewidth]{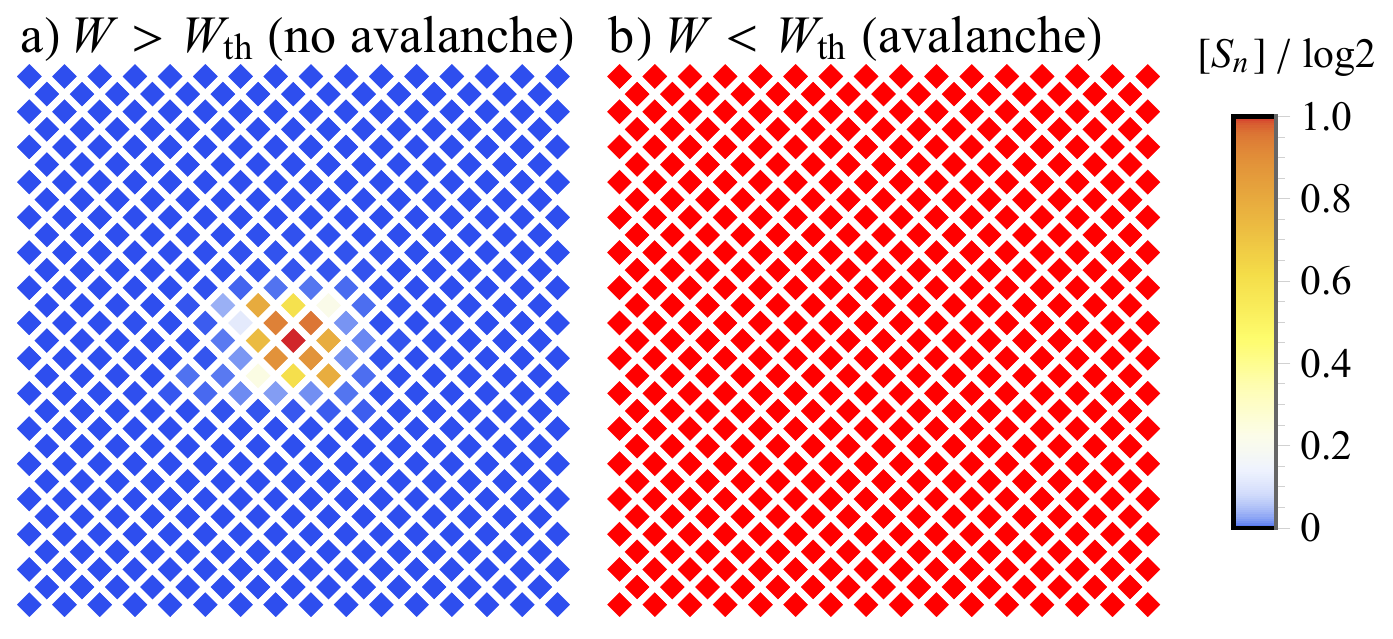}
    \caption{
    \emph{Avalanches and failed avalanches in 2D:} Orbitals on a square lattice (lattice vectors at $\pm 45^\circ$ from vertical) within lattice radius $R=15$ of the origin are shown. An ergodic grain is coupled at the origin. Orbital coloring denotes the state averaged entanglement entropy $[S_{\vec{n}}]$ (legend on right). In (a) the sample does not avalanche, and only orbitals close to the ergodic grain become thermal. In (b) the avalanche thermalizes all orbitals. Parameters: $W = 45 (20)$ left (right), $R=15$, $\theta_0 = 2.954$, $g_0 = 1$.
    }
    \label{Fig:2D1}
\end{figure}

 \begin{figure}[t!]
    \centering
    \includegraphics[width=\linewidth]{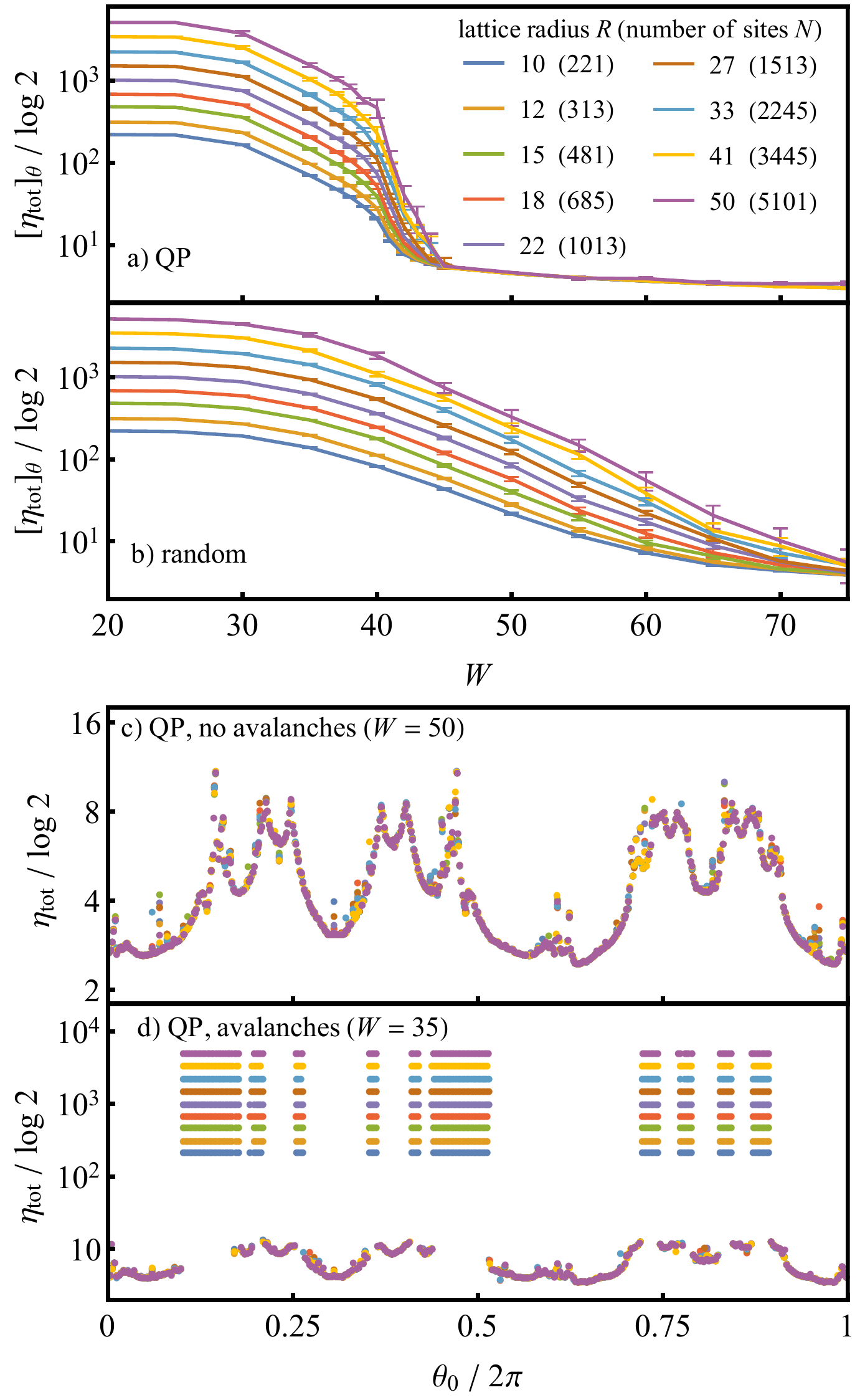}
    \caption{
    \emph{Avalanches in QP and random 2D systems:} Upper panels: the ensemble averaged total enhancement to the bubble entropy $[\eta_\mathrm{tot}]_\theta$ is plotted as a function of the potential strength $W$ for systems of different lattice radius (legend inset; number in brackets is $N$, the total number of sites) for (a) quasiperiodic and (b) iid random disorder. Lower panels: For the quasiperiodic case $\eta_\mathrm{tot}$ may be resolved as a function of $\theta_0$. For strong disorder (c) $\eta_\mathrm{tot}$ is continuous and converges in the limit of large $N$. For weaker disorder (d) the system avalanches for some values of $\theta_0$, for which $\eta_\mathrm{tot}$ scales as $N$. Fixed parameters: $g_{0} = 1$.
    }
    \label{Fig:2D2}
\end{figure}

In Fig.~\ref{Fig:2D1} we plot the l-bit entanglement entropies in a system of radius $R = 15$ for two cases. In Fig.~\ref{Fig:2D1}a the avalanche fails and only orbitals close to the ergodic grain thermalize. In Fig.~\ref{Fig:2D1}b the avalanche thermalizes all orbitals. These plots are obtained by solving the mean field equations with bare reduced couplings corresponding to~\eqref{eq:2D_cgm_diagform}
\begin{equation}
    g_{n,0} = g_0 |\phi_{\n\vec{0}}|
    \label{eq:2D_bare_couplings}
\end{equation}
where $g_0$ characterises the properties of the central grain. 

The mean field analysis exhibits a striking difference between QP and random potentials that has been frequently conjectured: for QP modulation the potential can be sufficiently strong that the system never avalanches. We see this in Figs.~\ref{Fig:2D2}a,~\ref{Fig:2D2}b where the ensemble averaged (i.e. average over $\theta_0$ for QP and $\theta_\n$ for random) total entropic enhancement $\eta_\mathrm{tot}$~\eqref{eq:eta_tot}, calculated for bare reduced couplings~\eqref{eq:2D_bare_couplings} for $g_0 = 1$. The enhancement satisfies the asymptotic equality
\begin{equation}
    [\eta_\mathrm{tot}]_{\theta} \sim N f_\mathrm{aval.} \log 2 
    \label{eq:faval}
\end{equation}
where $f_\mathrm{aval.}$ is the fraction of samples in the QP/random ensemble which avalanche. For QP potentials (Fig.~\ref{Fig:2D2}a) there are three regimes. At the smallest $W$, we find $[\eta_\mathrm{tot}]_{\theta} = N \log 2$, indicating that the system avalanches for all realizations of the potential, i.e. $f_\mathrm{aval.} = 1$. At stronger modulation there is a regime where $[\eta_\mathrm{tot}]_{\theta} = O(N) < N \log 2$, indicating the system avalanches for a finite fraction of realizations $0 < f_\mathrm{aval.} < 1$. Finally, at disorders above a finite ($g_0$ dependent) threshold value $W > W_\mathrm{th.} \approx 45$, we find $[\eta_\mathrm{tot}]_{\theta} = O(N^0)$ indicating that the system does not avalanche for any realizations~\footnote{Strictly $[\eta_\mathrm{tot}]_{\theta} = O(N^0)$ along implies that the system avalanches for at most a fraction $f_\mathrm{aval.} = O(1/N)$ of realizations---however, as avalanching is dictated by local physics, the fraction of avalanching samples cannot be $N$ dependent, so $f_\mathrm{aval.} = 0$, and avalanching samples are at most zero measure. Furthermore, as the disorder is continuous in $\theta_0$, and avalanching dictated by local physics, for any $\theta_0$ that avalanches, there will be a finite radius of nearby $\theta_0$ for which the system also avalanches, precluding the possibility of a measure zero set of avalanching samples.}. In contrast, for random potentials (Fig.~\ref{Fig:2D2}b), we find the data consistent with avalanching at all disorder strengths (i.e. $f_\mathrm{aval.} > 0$) for a fraction $f_\mathrm{aval.}$ which is monotonically decreasing in $W$ (indeed, for all $W$, $L$ analysed, we encountered avalanching samples). Indeed, simple arguments tell us this must be the case: uncorrelated random disorder always yields a finite probability of the disorder being uncharacteristically low in the vicinity of the grain, allowing the avalanche to reach the asymptotic regime where it may self sustain. This contrasts with the QP case, where varying the ensemble realization, i.e. $\theta_0$ (or equivalently the site to which the grain is coupled) does not lead to significant variation in the apparent potential strength.

In Figs.~\ref{Fig:2D2}c,~\ref{Fig:2D2}d we resolve $\eta_\mathrm{tot}$ as a function of $\theta_0$ for $W<W_\mathrm{th}$ and $W>W_\mathrm{th}$ respectively. Fig.~\ref{Fig:2D2}c shows that for $W > W_\mathrm{th.}$, for all system sizes, and for all $\theta_0$ the avalanche fails. Specifically, we find $\eta_\mathrm{tot}$ converges in the limit of $N \to \infty$ where the failed avalanche becomes insensitive to the system's boundary. The form of $\eta_\mathrm{tot}$ is continuous, though not smooth, at points the variation is so rapid that finite sampling leads to apparent discontinuities. In contrast, for $W < W_\mathrm{th.}$, shown in Fig.~\ref{Fig:2D2}d, different values of $\theta_0$ lead to different behaviours. For certain ranges of $\theta_0$  (corresponding approximately to regions of smaller $\eta_\mathrm{tot}$ in Fig.~\ref{Fig:2D2}c) the system does not avalanche and $\eta_\mathrm{tot} = O(N^0)$. For other ranges, the avalanche thermalizes the entire system, and $\eta_\mathrm{tot} = N \log 2$. 

Fig.~\ref{Fig:2D2}d further highlights a distinction between avalanches in $D=1$ and $D>1$. In QP systems, in $D>1$, avalanches cannot fail at arbitrarily large sizes. Specifically, for the parameters shown, we see that either avalanches succeed, yielding $\eta_\mathrm{tot} = N \log 2$, or fail, yielding shown $\eta_\mathrm{tot} \lesssim 14 \log 2$, with no possibility of failure at intermediate values. This observation justifies~\eqref{eq:faval}. We contrast this with the $D=1$ case (Fig.~\ref{Fig:DRH2}) where the avalanche may fail at an arbitrarily large size in the vicinity of the critical point~\cite{crowley2020avalanche}.

\paragraph*{\textbf{Stability of QP-MBL to avalanches in 2D}:} Consider introducing a finite density of grains into the QP system~\eqref{eq:2D_cgm}. Each grain has a slightly different local environment, parameterized by $\theta_0$. The failure of avalanches for all $\theta_0$ in Fig.~\ref{Fig:2D2}c suggests that this density of grains may not destabilize localization if the density is sufficiently small. Precisely, localization is stable provided $(i)$ each grain does not exceed a bounded initial size, and $(ii)$ each grain is sufficiently far from the others that the thermal bubbles do not merge---both conditions are natural for a quasiperiodic model. 

The mean field theory allows us to construct a conservative, but quantitative, estimate for the necessary spacing between grains. Specifically, we calculate the radius $R$ beyond which all l-bits experience only perturbative corrections, in all eigenstates, due to the failed avalanche. If all grains are separated by at least $2R$, the collars of non-perturbative influence do not overlap, and the failure of each avalanche is described by the mean field theory~\eqref{eq:MFeqs_N}. Consider a failed avalanche in which the total entropic enhancement of the grain is $\eta_\mathrm{tot}$. The thermal bubble has a dimension $d = d_\mathrm{g} \e^{\eta_\mathrm{tot}}$, and the reduced coupling of the $\vec{n}$th l-bit to the bubble is $g_{\vec{n}} = g_{\vec{n},0} \e^{\eta_\mathrm{tot}/2}$. The $n$th l-bit is resonant in $O(g_{\vec{n}} d)$ states, and only perturbatively corrected in others. Thus, in the weak coupling regime $g_{\vec{n}} \ll 1/d$, the $\vec{n}$th l-bit is perturbatively corrected in all eigenstates for a typical disorder realization. By extension, the total number of resonances involving any l-bits outside the radius $R$ is $O(\sum_{\n : R < |\n|} g_{\vec{n}} d)$. Hence, within a typical disorder realization~\footnote{We note the possibility of $O(1)$ many-body resonances involving l-bits outside this collar is inconsequential for the stability of localization, indeed such many-body resonances are a generic feature of the MBL phase~\cite{crowley2020constructive,garratt2021local}}, all such l-bits experience only perturbative corrections in all eigenstates if
\begin{equation}
    \sum_{\n : R < |\n|} g_\n \ll 1/d.
    \label{eq:R}
\end{equation}
Eq.~\eqref{eq:R} may be used to define $R$. As $g_\n$ is exponentially small in $|\n|,$ solutions are generic. 

In the model~\eqref{eq:2D_cgm}, we have considered the effect of interactions only within the ergodic grains. In contrast, generic models of MBL have interactions everywhere. Nevertheless, at strong modulation, the effect of interactions is typically perturbative. If one identifies spatial regions where the effects of interactions are non-perturbative with the ergodic grains, then~\eqref{eq:2D_cgm} provides a toy model for their influence on surrounding l-bits. We then expect that avalanches do not destabilize generic QP-MBL. 

However, we note that the toy model does not account for two features numerically observed in generic MBL. Consider first the ``Hartree shifts'' to the energies of the orbital configurations, e.g. terms of the form $f_\n^\dagger f_\n f_\m^\dagger f_\m$.  These shifts result in small changes to the decoupled many-body energies, and thus de-tune certain resonances, whilst bringing other pairs of states into resonance. Overall, the statistics of resonances across eigenstates (in particular the reduced coupling $g_n$) is unchanged. Second, the toy model~\eqref{eq:2D_cgm} neglects couplings that act on the grain and multiple orbitals simultaneously, e.g. $c_\mathrm{g}^\dagger f_\n^\dagger f_\m f_{\vec{p}}$. A particular class couple specific pairs of distinct orbital configurations, e.g. $(\ket{E_a}\bra{E_b} + \mathrm{h.c})$. For rare pairs, such terms are uncharacteristically large, corresponding to when $\ket{E_a}$, $\ket{E_b}$ are many-body resonances in the lattice basis~\cite{gopalakrishnan2015low,crowley2020constructive,garratt2021local,morningstar2021avalanches}. Nevertheless, we expect multi-orbital terms which involve rearrangements in a large radius are sufficiently suppressed so that only the collar region of the failed avalanche is quantitatively modified.

\paragraph*{\textbf{Discussion}:} 
We have developed a self-consistent \emph{entanglement mean-field theory} of central grain models. In this theory, the coupling between l-bits and the central grain leads to many-body resonances between the eigenstates as calculated for zero coupling. We quantify these resonances in terms of the reduced couplings $g_n$~\eqref{eq:pg1a}. The reduced coupling parameterizes the distribution of hybridization strengths~\eqref{eq:g1a}, closely related quantities are studied in Refs~\cite{serbyn2015criterion,thiery2018many,maksymov2019energy,crowley2020avalanche,sels2020dynamical}. At small $g_n$, the reduced coupling is equal to the fraction of eigenstates in which the $n$th l-bit is resonant (i.e. non-perturbatively hybridized)~\cite{crowley2022partial}. We describe how $g_n$ may be self-consistently calculated, and used to determine the infinite time properties of the system: namely, the distribution of eigenstate entanglement entropies, and the infinite time memory of observables. 

The mean-field theory describes l-bit properties in eigenstates at maximum entropy in a central grain model. We leave to future work the extension of this theory to different geometries; finite temperature effects; systems in which the l-bits have more than two levels; and to include multi-l-bit couplings.

The mean-field theory quantitatively captures how the process of resonance formation may run away leading to an avalanche, or conversely how the avalanche may fail, resulting in l-bits which are partially thermalized, having a bi-modal distribution of entanglement entropies across eigenstates (see Fig.~\ref{Fig:1}c and Fig.~\ref{Fig:DRH1}).
Beyond the central spin geometry, several recent works have explored the role of many-body resonances in many-body delocalization~\cite{gopalakrishnan2015low,crowley2020constructive,villalonga2020eigenstates,garratt2021local,tikhonov2021eigenstate,tikhonov2021anderson,morningstar2021avalanches}. 

Correlations in a localizing potential can alter the dynamical properties of a system~\cite{luck1993critical,luck1993classification,crowley2018quasiperiodic,crowley2018critical,chandran2017localization,crowley2019quantum,cookmeyer2020critical}. In particular, while sufficiently small higher dimensional systems may appear many-body localized~\cite{kennes2018many,wahl2019signatures,theveniaut2020transition,chertkov2021numerical,decker2021many,devakul2017anderson,choi2016exploring,bordia2017probing}, it is argued that for random potentials avalanches destabilize MBL~\cite{de2017stability}. In contrast, MBL due to quasiperiodic potentials has long been of interest~\cite{iyer2013many,vznidarivc2018interaction,mace2019many,varma2019diffusive,doggen2019many,singh2021local,aramthottil2021finite,vstrkalj2021many,zhang2018universal}, due to the conjecture that QP-MBL may not suffer from the avalanche stability, altering the universality class of the 1D MBL-thermal transition~\cite{khemani2017two}, and stabilizing the phase in 2D.

We provide analytic evidence that a grain in a two-dimensional random potential has a finite probability of inducing an avalanche at any modulation strength, unlike in the QP case, where for a sufficiently strong potential, the system never avalanches. On this basis, we argue for the stability of QP-MBL to avalanches in 2D. However, we note that the stability of MBL to other destabilizing processes has recently been a subject of active debate~\cite{vsuntajs2020quantum,sels2020dynamical,crowley2020constructive,sels2021markovian,abanin2021distinguishing,leblond2021universality,morningstar2021avalanches,garratt2021local}. 

A further conceptual insight provided by mean field theory is that a weak coupling to sufficiently many l-bits, as opposed to a sufficiently strong coupling to a single l-bit, can sustain an avalanche. This is illustrated most simply in the central grain model with symmetric couplings $g_{n,0} = g_0$, corresponding closely to the models of Refs.~\cite{ponte2017thermal,hetterich2018detection,ng2019many,ashida2019quantum}. In this case the mean field equations are correspondingly symmetric, with $g_n = g$ given by
\begin{equation}
    g = g_0 \e^{N \eta(g)/2}
    \label{eq:MF_cspin}
\end{equation}
By straightforward analysis of~\eqref{eq:MF_cspin} (using the asymptotic relation $\eta(g) \sim - 8 g \log g$), the enhancement entropy is non-extensive, $\eta_\mathrm{tot} = N \eta(g) \sim N^0$, and hence the system is localized, only for
\begin{equation}
    g_0 < g_\mathrm{c} \sim (4 N \log N)^{-1}
    \label{eq:MF_cspin_gc}.
\end{equation}
We note this critical value agrees with the breakdown of localization predicted by Ref.~\cite{altshuler1997quasiparticle} (though the mean field theory does not produce the non-ergodic delocalized phase reported between $g_\mathrm{c} \propto (N \log N)^{-1}$ and $g_\mathrm{c}' \propto N^{-1}$~\cite{altshuler1997quasiparticle,ponte2017thermal}). Thus, for arbitrarily weak couplings coupling $g_0$, one can always increase $N$ to violate~\eqref{eq:MF_cspin_gc} and  thermalize all l-bits.

\paragraph*{\textbf{Acknowledgements}:} We are grateful to C.R. Laumann and A. Polkovnikov for useful discussions, and to S. Garratt, D. Huse and V. Oganesyan for useful comments. P.C. is supported by the NSF STC ``Center for Integrated Quantum Materials'' under Cooperative Agreement No. DMR-1231319. A.C is supported by NSF DMR-1813499. Numerics were performed on the Shared Computing Cluster, administered by Boston University’s Research Computing Services.

\paragraph*{\textbf{Note added}:} During preparation of this manuscript, we became aware of Refs.~\cite{agrawal2022note} and~\cite{strkalj2022coexistence}. Ref.~\cite{agrawal2022note} argues for the stability of 2D QP-MBL phase against avalanches from a complementary perspective, reaching conclusions in agreement with this manuscript. Ref.~\cite{strkalj2022coexistence} argues for the stability of 2D QP-MBL even in the presence of infinite ergodic regions, a claim which is inconsistent with our results.

\bibliography{bib}

\appendix

\section{Enhancement to the bath due to intermediately coupled spins}

\label{app:eta}

In this section, we recap and extend the calculation of Ref.~\cite{crowley2022partial}. We derive the entropic enhancement to the ergodic grain due to multiple intermediately coupled spins. For simplicity we assume $H_\mathrm{g}$ to be a GOE matrix, however the results are readily generalizable to the case of an ETH satisfying system (see Ref.~\cite{crowley2022partial}). The results are obtained using an ansatz for the many body eigenstates which accounts for the effect of resonances. The main results of this appendix are:
\begin{enumerate}
    \item For an $N=2$ spin central grain model, where $g_{1,0}$ and $g_{2,0}$ are in the intermediate and weak coupling regimes respectively, the reduced coupling of the second spin is given by 
    \begin{equation}
        g_{2} = g_{2,0} \exp\left(\tfrac12 \eta(g_1)\right)
    \end{equation}
    where $g_1 = g_{1,0}$, and the entropic enhancement $\eta(g)$ is given by~\eqref{eq:eta_app}.
    \item For the general case of $N$ intermediately coupled spins, and a single $(N+1)$th spin in the weak coupling regime, the reduced coupling of the weakly coupled spin is enhanced by an additive entropic term
    \begin{equation}
        g_{N+1} = g_{N+1,0} \exp\left(\tfrac12 \sum_{n=1}^N\eta(g_n)\right).
    \end{equation}
    where, as before, $\eta(g)$ is given by~\eqref{eq:eta_app}.
\end{enumerate}
The form of $\eta(g)$ derived in this appendix~\eqref{eq:eta_app} is used in numerical calculations throughout this manuscript, and plotted in Fig~\ref{Fig:1}. 

\subsection{Single spin}

We begin by recapitulating the properties of the many body eigenstates in the $N=1$ case, studied in detail in Ref.~\cite{crowley2022partial}. 

When spin is in the intermediate coupling regime, the infinite time properties (i.e. distribution of entanglement entropies and time averaged autocorrelators) of the spin are determined by the distribution of the quantity $g_{1a}$---the norm of the first corrections in perturbation theory as $V_1$ is introduced~\eqref{eq:g1a}. As in the main text, we denote the distribution of this quantity with $p_{g_{1a}}(g)$ so that the ensemble averaged fraction of the $g_{1a}$ in the interval $[g , g + \d g]$ is given by $p_{g_{1a}}(g) \d g$. For the GOE grain considered here $p_{g_{1a}}(g)$ may be calculated
\begin{equation}
    \begin{aligned}
        p_{g_{1a}}(g) &= \tilde{p}(g/g_1)/g_1
        \\
        \tilde{p}(x) & = \frac{2}{x^2} \exp \left(- \frac{\pi^3}{4x^2}\right) \bigg( 1 + \frac{c_1}{x} + \frac{c_2}{x^2}  +O\left(x^{-3}\right)\bigg)
    \end{aligned}
    \label{eq:p_goe}
\end{equation}
where $c_1 = 5.3\ldots$, $c_2 = 11.2\ldots$. However, the main features are more general: (i) a rapid decay for $g \lesssim g_1$ (ii) a unimodal peak around the typical value $g \approx g_1$ and (iii) a power law tail at large $g$ given asymptotically by
\begin{equation}
    p_{g_{1a}}(g) \sim 2g_1/g^2.
\end{equation}
Here $g_1$ is defined (as in~\eqref{eq:g1}) by
\begin{equation}
    g_1 = \rho [|V_{1,ab}|].
\end{equation}
where $\rho$ is the many body density of states at maximum entropy, and $V_{1,ab}$ are the matrix elements of $V_1$ between the eigenstates $\ket{E_a}$ of $H$ evaluated for $V_1 = 0$, and the square brackets $[\cdot]$ denotes averaging over $a \neq b$. Each of these eigenstates is a product state of the first spin and grain
\begin{equation}
    \ket{E_a} = \ket{\epsilon_\alpha}\ket{\sigma}
    \label{eq:app_eig}
\end{equation}
for $a = (\alpha,\sigma)$, $\sigma \in \{ \up , \dn \}$, and $\ket{\epsilon_\alpha}$ an eigenstate of the grain $H_\mathrm{g}$.

Consider following the eigenstates as the coupling $V_1$ tuned from zero to its finite value. As the spin is in the intermediate coupling regime, typically eigenstates are only perturbatively altered. However, a minority of states $\ket{E_a}$ are accidentally close to states from the opposite spin sector, and have correspondingly large values $g_{1a} \gtrsim 1$. These states strongly hybridize, forming resonances (depicted in Fig.~\ref{Fig:1}b). Typically, such resonances involve only a pair of states: the state $\ket{E_a}$, and some other nearby state $\ket{E_c}$. We may thus approximate $g_{1a}$
\begin{equation}
    g_{1a} = \sqrt{\sum_{b \neq a} \left| \frac{V_{1,ab}}{E_a - E_b} \right|^2 } \approx \left| \frac{V_{1,ac}}{E_a - E_c} \right|
    \label{eq:g1a_approx}
\end{equation}
where $c$ is obtained by minimising the denominator $|E_a - E_b|$.

An ansatz for the eigenstates is obtained by diagonalizing within these two-level resonance sub-spaces. The effective two-level Hamiltonian is given by
\begin{equation}
    H_\mathrm{eff} = \begin{pmatrix}
    E_a & V_{1,ac}
    \\
    V_{1,ac} & E_c
    \end{pmatrix} 
    \approx
    \Delta_{ac} \begin{pmatrix}
    1 & g_{1a}
    \\
    g_{1a} & 0
    \end{pmatrix} + E_c
    \label{eq:Heff}
\end{equation}
where $\Delta_{ac} = E_a - E_c$ and we have used~\eqref{eq:g1a_approx} that $g_{1a} \approx | V_{1,ac}/\Delta_{ac} |$. Explicit diagonalization of~\eqref{eq:Heff} yields the new eigenvectors
\begin{equation}
    \begin{aligned}
    \ket{E_a'} & = \sqrt{Q_{a}} \ket{E_a} + \sqrt{P_{a}} \ket{E_c}
    \\
    \ket{E_c'} & = \sqrt{P_{a}} \ket{E_a} - \sqrt{Q_{a}} \ket{E_c}
    \end{aligned}
    \label{eq:eig_1_spin}
\end{equation}
where $P_a = P(g_{1a})$, $Q_a = Q(g_{1a})$ are given by
\begin{equation}
    P(g) := 1 - Q(g) := \frac12 \left( 1 - \frac{1}{\sqrt{1 + 4 g^2}} \right).
    \label{eq:pq_def}
\end{equation}

\subsubsection*{Infinite time observables}
\label{app:inf_time}

This eigenstate ansatz~\eqref{eq:eig_1_spin} can be used to calculate the distribution of eigenstate entanglement entropies and infinite time memory of the spin, and agrees with numerics~\cite{crowley2022partial}. 

Specifically, we have that the entanglement entropy $S_{na}$ of the $(n=1)$th spin the $a$th state is given by $S_{na} = S(g_{na})$ with
\begin{equation}
    \begin{aligned}
    S(g) & = - P(g) \log P(g) - Q(g) \log Q(g)
    \end{aligned}
\end{equation}
and thus follows a distribution
\begin{equation}
    p_{S_{na}}(S_n) = \int \d g \delta(S_n - S(g)) p_{g_{na}}(g)
\end{equation}
with mean value
\begin{equation}
    [S] = \int \d g S(g) p_{g_{na}}(g).
\end{equation}
By similar arguments the infinite time correlator is given by
\begin{equation}
    \overline{C_n^{zz}} = \int \d g p_{g_{na}}(g) (P(g)-Q(g))^2.
\end{equation}
In the intermediate regime it is sufficient to write
\begin{equation}
    p_{g_{na}}(g) = g_n / g^2 + O(g_n^2/g^3)
\end{equation}
to obtain the limits given in the main text~\eqref{eq:S_Czz} (details in Ref.~\cite{crowley2022partial}).

\subsection{Two spins}

We next consider introducing a second spin. Again, this case was considered in Ref.~\cite{crowley2022partial}, and we here recap the calculation. The `effective bath' seen by the second spin is composed of the first spin and grain. As a result, the reduced coupling is enhanced from its bare ($V_1=0$) value
\begin{equation}
    g_{2} = g_{2,0} \exp\left(\tfrac12 \eta(g_1)\right)
    \label{eq:g2_app}
\end{equation}
in this section we calculate $\eta(g)$.

We begin by considering $p_{g_{2a}}(g)$, the distribution of $g_{2a}$ the norm of the first order term in perturbation theory when the eigenstate $\ket{E_a}$ is expanded about $V_2 = 0$, but for $V_1$ finite. This distribution has the same qualitative features as $p_{g_{1a}}(g)$: (i) a rapid decay for $g \lesssim g_2$ (ii) a unimodal peak around the typical value $g \approx g_2$ and (iii) a power law tail at large $g$ given asymptotically by
\begin{equation}
    p_{g_{2a}}(g) \sim 2g_2/g^2
\end{equation}
where 
\begin{equation}
    g_2 = \rho [|V_{2,ab}'|].
    \label{eq:g2_app2}
\end{equation}
where $V_{2,ab}'$ are the matrix elements of $V_2$ between the eigenstates $\ket{E_a'}$ of $H$ evaluated for $V_2 = 0$ but $V_1$ finite. Each of these eigenstates is a product state of the second spin and grain, but in general is an entangled state of the first spin and grain~\eqref{eq:eig_1_spin}. From~\eqref{eq:g2_app} and~\eqref{eq:g2_app2} it follows that
\begin{equation}
    \eta(g_1) = 2 \log \left(\frac{{[|{V_{2,ab}^{\mathrm{\prime}}}|]}}{[|V_{2,ab}|] }\right)
    \label{eq:eta_app2}
\end{equation}
where $V_{2,ab}$ are the matrix elements of $V_2$ between the eigenstates $\ket{E_a}$ of $H$ evaluated for $V_2 = 0$ and $V_1 = 0$. In the remainder of this section, we evaluate~\eqref{eq:eta_app2}.

Each many body eigenstate may be identified with a sector of the first spin by following the state adiabatically as we tune $V_1 \to 0$, and measuring the state of the first spin. Using this labelling scheme, we see that there are two qualitatively different families of matrix elements: $V_{2,ab}'$ is \emph{even} if $\ket{E_a'}$ and $\ket{E_b'}$ correspond to the same sector, and \emph{odd} otherwise. Half of the matrix elements are even, and half odd, so that
\begin{equation}
    [|V_{2,ab}'|] = \frac12 \left( [|{V_{2,ab}^{\mathrm{(e)}\prime}}|] + [|{V_{2,ab}^{\mathrm{(o)}\prime}}|] \right).
    \label{eq:VOE}
\end{equation}
where $[|{V_{2,ab}^{\mathrm{(e)}\prime}}|]$, $[|{V_{2,ab}^{\mathrm{(o)}\prime}}|]$ are the averaged taken over the odd/even sectors only. We evaluate each of these contributions in turn.

We first evaluate the even sector. We consider two generic states
\begin{subequations}
    \begin{align}
    \ket{E_a'} & 
    = \sqrt{Q_{a}} \ket{E_a} + \sqrt{P_{a}} \ket{E_c} 
    \nonumber
    \\
    & = \sqrt{Q_{a}} \ket{\epsilon_\alpha}\ket{\up_1} + \sqrt{P_{a}} \ket{\epsilon_\beta}\ket{\dn_1}
    \label{eq:1spin_eig}
    \\
    \ket{E_b'} & 
    = \sqrt{Q_{b}} \ket{E_b} + \sqrt{P_{b}} \ket{E_d}
    \nonumber
    \\
    & = \sqrt{Q_{b}} \ket{\epsilon_\gamma}\ket{\up_1} + \sqrt{P_{b}} \ket{\epsilon_\delta}\ket{\dn_1}
    \end{align}
\end{subequations}
where in each case in the second line we have denoted the state of the first spin $\sigma_1$ explicitly, following~\eqref{eq:app_eig}. As $V_2$ does not act on the first spin (i.e. $[V_2,\sigma_1^\alpha]=0$), the corresponding matrix element contains only two terms
\begin{equation}
    \begin{aligned}
        {V_{2,ab}^{\mathrm{(e)}\prime}} & = \sqrt{Q_a Q_b} V_{2,ab}^{\mathrm{(e)}} + \sqrt{P_a P_b} V_{2,cd}^{\mathrm{(e)}} 
    \end{aligned}
\end{equation}
where $V_{2,ab}^{\mathrm{(e)}}$ are the even matrix elements of $V_2$ calculated in the basis $\ket{E_a}$, the eigenbasis calculated for both $V_2 = 0$ \emph{and} $V_1 = 0$. Note that by this definition the odd elements are zero $V_{2,ab}^{\mathrm{(o)}} = 0$.

As the $P_a = P(g_{1a})$ describe resonances induced by $V_1$, they are uncorrelated with the matrix elements $V_{2,ab}$. Moreover, as the bare grain is ergodic, the matrix elements $V_{2,ab}$, $V_{2,cd}$ are iid Gaussian distributed. For iid Gaussian distributed random variables $x,x'$ with zero mean $[x] = [x']=0$, it follows have that $[|a x + b x'|] = \sqrt{a^2 + b^2}[|x|]$. Using this, we obtain
\begin{equation}
    \begin{aligned}
    {[|{V_{2,ab}^{\mathrm{(e)}\prime}}|]} & = [|V_{2,ab}^{\mathrm{(e)}}|] [\sqrt{P_a P_b + Q_a Q_b}]
    \\
    & = 2[|V_{2,ab}|] [\sqrt{P_a P_b + Q_a Q_b}].
    \end{aligned}
\end{equation}
where in the second line we have used that $[|V_{2,ab}|] = \tfrac12 ( [|{V_{2,ab}^{\mathrm{(e)}}}|] + [|{V_{2,ab}^{\mathrm{(o)}}}|] ) = \tfrac12 [|{V_{2,ab}^{\mathrm{(e)}}}|] $. Furthermore, $P_a$ and $P_b$ are uncorrelated, so we obtain
\begin{equation}
    \begin{aligned}
    {[|{V_{2,ab}^{\mathrm{(e)}\prime}}|]} & = 2[|V_{2,ab}|] \iint \d g \d g' p_{g_{1a}}(g) p_{g_{1a}}(g') K^{\mathrm{(e)}} (g,g')
    \\
    K^{\mathrm{(e)}} (g,g') & =  \sqrt{P(g) P(g')+ Q(g) Q(g')} 
    \end{aligned}
    \label{eq:Ke}
\end{equation}
Repeating the same series of arguments for the odd terms we obtain
\begin{equation}
    \begin{aligned}
    {[|{V_{2,ab}^{\mathrm{(o)}\prime}}|]} & = 2[|V_{2,ab}|] \iint \d g \d g' p_{g_{1a}}(g) p_{g_{1a}}(g') K^{\mathrm{(o)}} (g,g')
    \\
    K^{\mathrm{(o)}} (g,g') & =  \sqrt{P(g) Q(g')+Q(g) P(g')} 
    \end{aligned}
    \label{eq:Ko}
\end{equation}
By substituting~\eqref{eq:Ke} and~\eqref{eq:Ko} into~\eqref{eq:VOE} we have
\begin{equation}
    \begin{aligned}
    {[|{V_{2,ab}^{\mathrm{\prime}}}|]} & = [|V_{2,ab}|] \iint \d g \d g' p_{g_{1a}}(g) p_{g_{1a}}(g') K (g,g')
    \\
    K (g,g') & = K^{\mathrm{(e)}} (g,g') + K^{\mathrm{(o)}} (g,g')
    \end{aligned}
    \label{eq:V2prime}
\end{equation}
Substituting~\eqref{eq:V2prime} into~\eqref{eq:eta_app} we obtain the entropic enhancement to the effective bath due to hybridization between the grain and first spin
\begin{equation}
    \eta(g) = 2 \log\left( \iint \d g \d g' p_{g_{1a}}(g) p_{g_{1a}}(g') K (g,g') \right).
    \label{eq:eta_app}
\end{equation}
where the right-hand side depends on $g_1$ via its appearance in $p_{g_{1a}}$~\eqref{eq:p_goe}.

\subsection{Three spins}

We next consider introducing a third spin. To calculate $g_3$ we require a model for the eigenstates of the central grain model for finite $V_1$ and finite $V_2$. We generalize the ansatz~\eqref{eq:eig_1_spin} to the two spin case
\begin{equation}
    \begin{aligned}
    \ket{E_a''} & = \sqrt{Q_{a} Q_{a}'} \ket{E_a} + \sqrt{P_{a} Q_{a}'} \ket{E_b}
    \\
    & \quad 
    + \sqrt{Q_{a} P_{a}'}\ket{E_c} +
    \sqrt{P_{a} P_{a}'} \ket{E_d}
    \end{aligned}
    \label{eq:2spin_eig}
\end{equation}
where $P_{a} = 1 - Q_{a} = P(g_{1a})$ and $P_{a}' = 1 - Q_{a}' = P(g_{2a})$ $\sigma_n \in \{ \up, \dn \}$ and we use a bar notation to denote spin flips so that $\bar{\up} = \dn$, $\bar{\dn} = \up$, and the $\ket{E_a}, \cdots , \ket{E_d}$ are eigenstates in the limit $V_1,V_2\to 0$, i.e. product states of the two spins
\begin{equation}
    \begin{aligned}
    \ket{E_a} & = \ket{\epsilon_\alpha}\ket{\sigma_1}\ket{\sigma_2},
    & 
    \ket{E_b} & =
    \ket{\epsilon_\beta}\ket{\bar{\sigma}_1}\ket{\sigma_2} ,
    \\ 
    \ket{E_c} & = \ket{\epsilon_\gamma}\ket{\sigma_1}\ket{\bar{\sigma}_2},
    &
    \ket{E_d} & = \ket{\epsilon_\delta}\ket{\bar{\sigma}_1}\ket{\bar{\sigma}_2}.
    \end{aligned}
\end{equation}

The calculation of $[|V_{3,ab}''|]$ then proceeds in direct generalization of the previous section. However, now there are more `species' of matrix element. Consider adiabatically following the eigenstates as we take the limit of $g_1,g_2 \to 0$: the state~\eqref{eq:2spin_eig} tends to the  $(\sigma_1,\sigma_2)$ sector of the two spins. As before, we use a convention in which we associate the state~\eqref{eq:2spin_eig} with the $(\sigma_1,\sigma_2)$ sector even for finite $V_1$, $V_2$. Consequently, there are four species of matrix element $V_{3,ab}''$ depending on whether $\sigma_1$, $\sigma_2$, both or neither must changed to relate the states $\ket{E_a''}$ and $\ket{E_b''}$, which we refer to as the (odd,even), (even,odd), (odd,odd) and (even,even) sectors respectively. 

The mean matrix element is obtained by averaging across these sectors
\begin{equation}
    \begin{aligned}
    {[|V_{3,ab}''|]} = & \frac14 \left( [|{V_{3,ab}^{\mathrm{(e,e)}\prime\prime}}|] +
    [|{V_{3,ab}^{\mathrm{(e,o)}\prime\prime}}|]  \right.
    \\
    & \quad \left. +
    [|{V_{3,ab}^{\mathrm{(o,e)}\prime\prime}}|] +
    [|{V_{3,ab}^{\mathrm{(o,o)}\prime\prime}}|] \right).
    \label{eq:VOOEE}
    \end{aligned}
\end{equation}
These are calculated following the same prescription of the previous section. In the (even,even) case we consider the matrix element between two states, e.g. $\ket{E_a''}$ from~\eqref{eq:2spin_eig} and 
\begin{equation}
    \begin{aligned}
    \ket{E_e''} & = \sqrt{Q_{a} Q_{a}'} \ket{E_e} + \sqrt{P_{a} Q_{a}'} \ket{E_f}
    \\
    & \quad 
    + \sqrt{Q_{a} P_{a}'}\ket{E_g} +
    \sqrt{P_{a} P_{a}'} \ket{E_h}
    \end{aligned}
\end{equation}
where similarly
\begin{equation}
    \begin{aligned}
    \ket{E_e} & = \ket{\epsilon_\epsilon}\ket{\sigma_1}\ket{\sigma_2},
    & 
    \ket{E_f} & =
    \ket{\epsilon_\zeta}\ket{\bar{\sigma}_1}\ket{\sigma_2} ,
    \\ 
    \ket{E_g} & = \ket{\epsilon_\eta}\ket{\sigma_1}\ket{\bar{\sigma}_2},
    &
    \ket{E_h} & = \ket{\epsilon_\theta}\ket{\bar{\sigma}_1}\ket{\bar{\sigma}_2}.
    \end{aligned}
\end{equation}
Computing the matrix element directly, we use that $V_3$ does not alter the state of the first or second spin, and obtain
\begin{widetext}
\begin{equation}
     {V_{3,ae}^{\mathrm{(e,e)}\prime\prime}} = \sqrt{Q_a Q_a ' Q_{e} Q_{e}'} \bra{E_a } V_3 \ket{ E_e } +  
     \sqrt{P_a Q_a ' P_{e} Q_{e}'}  \bra{E_b } V_3 \ket{ E_f } +
     \sqrt{Q_a P_a ' Q_{e} P_{e}'} \bra{E_c } V_3 \ket{ E_g } +
     \sqrt{P_a P_a ' P_{e} P_{e}'} \bra{E_d } V_3 \ket{ E_h } 
\end{equation}
with a mean size
\begin{align}
    {[|V_{3,ae}^{\mathrm{(e,e)}\prime\prime}|]} & = [|V_{3,ae}^{\mathrm{(e,e)}}|] [\sqrt{ P_a P_a ' P_{e} P_{e}' + Q_a P_a ' Q_{e} P_{e}' + P_a Q_a ' P_{e} Q_{e}' + Q_a Q_a ' Q_{e} Q_{e}' } ]
    \nonumber
    \\
    & = [|V_{3,ae}^{\mathrm{(e,e)}}|] [\sqrt{ P_a P_a ' + Q_a Q_a '} \sqrt{  P_{e} P_{e}' +  Q_{e} Q_{e}' } ]
    \nonumber
    \\
    & = 4 [|V_{3,ae}|] [\sqrt{ P_a P_a ' + Q_a Q_a '}][ \sqrt{  P_{e} P_{e}' +  Q_{e} Q_{e}' } ]
    \nonumber
    \\
    & = 4 [|V_{3,ae}|] \left(\iint \d g \d g' \ p_{g_{1a}}(g) p_{g_{1a}}(g') K^{\mathrm{(e)}} (g,g') \right) \left(\iint \d g \d g' \ p_{g_{2a}}(g) p_{g_{2a}}(g') K^{\mathrm{(e)}} (g,g') \right)
\end{align}
Repeating this calculation for the other sectors we obtain
\begin{equation}
    \begin{aligned}
    {[|V_{3,ae}^{(s,s')\prime\prime}|]} &  = 4 [|V_{3,ae}|] \left(\iint \d g \d g' \ p_{g_{1a}}(g) p_{g_{1a}}(g') K^{(s)} (g,g') \right) \left(\iint \d g \d g' \ p_{g_{2a}}(g) p_{g_{2a}}(g') K^{(s')} (g,g') \right)
    \end{aligned}
\end{equation}
for $s,s'\in \{ \mathrm{o},\mathrm{e}\}$ and hence, using~\eqref{eq:VOOEE}.
\begin{equation}
    \begin{aligned}
    {[|V_{3,ae}^{\prime\prime}|]} &  =  [|V_{3,ae}|] \left(\iint \d g \d g' \ p_{g_{1a}}(g) p_{g_{1a}}(g') K (g,g') \right) \left(\iint \d g \d g' \ p_{g_{2a}}(g) p_{g_{2a}}(g') K (g,g') \right)
    \end{aligned}
\end{equation}
\end{widetext}
With this result, we determine that the enhancement is given by a sum of the enhancements due to the two spins.
\begin{equation}
    \log \left( \frac{g_3}{g_{3,0}} \right) = \log \left(\frac{{[|{V_{3,ab}^{\mathrm{\prime\prime}}}|]}}{[|V_{3,ab}|] }\right) = \tfrac12 \left( \eta(g_1) + \eta(g_2) \right).
\end{equation}

\subsection{$N$ spins}

The many spin case is found by direct generalization of the previous section. Here we simply state the eigenstate ansatz, and the result. 

The ansatz for an eigenstate associated to the sector $\vec{\sigma} = (\sigma_1,\sigma_2 \cdots \sigma_N)$ is given by
\begin{equation}
    \begin{aligned}
    \ket{E_a^{(n)}} & = \sum_{\vec{\tau}} C_{\vec{\sigma},\vec{\tau}} \ket{\epsilon_{\alpha,\vec{\tau}}}\ket{\tau_1}\ket{\tau_2}\cdots \ket{\tau_N} 
    \end{aligned}
    \label{eq:Nspin_eig}
\end{equation}
where 
\begin{subequations}
\begin{align}
    C_{\vec{\sigma},\vec{\tau}} & = \prod_{n} c_{\sigma_n,\tau_n} 
    \\
    c_{\sigma_n,\tau_n} & = \begin{cases}
    \sqrt{ P(g_{na}) } \quad & \text{if} \quad \sigma_n = \tau_n
    \\
    \sqrt{ Q(g_{na}) } \quad & \text{if} \quad \sigma_n \neq \tau_n
    \end{cases}
\end{align}
\end{subequations}
and $\ket{\epsilon_{\alpha,\vec{\tau}}}$ are a set of distinct eigenvectors of the grain. The form~\eqref{eq:Nspin_eig} can be seen to reduce to the forms~\eqref{eq:1spin_eig} and~\eqref{eq:2spin_eig} in the cases $N=1$ and $N=2$. By generalizing the above calculation, we obtain 
\begin{equation}
    \log \left( \frac{g_{N+1}}{g_{N+1,0}} \right)  = \frac12 \sum_{n = 1}^N \eta(g_n) 
\end{equation}
as desired.

\end{document}